\documentclass{appolb}
\usepackage{epsfig}
\usepackage[tbtags]{amsmath}
\begin{document}
%\date{\today}
\pagestyle{plain}
%% uncomment the following line to get equations numbered by (sec.num)
%\eqsec
\newcount\eLiNe\eLiNe=\inputlineno\advance\eLiNe by -1
\title{THE EFFECT OF THE PAIRING INTERACTION ON THE ENERGIES OF
ISOBAR ANALOG RESONANCES IN $^{112-124}$Sb AND ISOSPIN ADMIXTURE
IN $^{100-124}$Sn ISOTOPES}

\author{Tahsin BABACAN$^*$, Djavad SALAMOV$^\dagger$, \\Atalay K{\"U}{\d{C}}{\"U}KBURSA$^\dagger$,
Halil BABACAN$^*$, Ismail MARA{\d{S}}$^*$, \\Hasan A.
AYG{\"O}R$^*$, Arslan {\"U}NAL$^\dagger$
\address{$^*$Department of Physics, Celal Bayar University,
Manisa-Turkey \\$^\dagger$Department of Physics, Dumlup{\i}nar
University, K{\"u}tahya-Turkey}} \maketitle

\begin{abstract}
In the present study,  the effect of the pairing interaction and
the isovector correlation between nucleons on the properties of
the isobar analog resonances (IAR) in $^{112-124}$Sb isotopes and
the isospin admixture in $^{100-124}$Sn isotopes is investigated
within the framework of the quasiparticle random phase
approximation (QRPA). The form of the interaction strength
parameter is related to the shell model potential by restoring the
isotopic invariance of the nuclear part of the total Hamiltonian.
In this respect, the isospin admixtures in the $^{100-124}$Sn
isotopes are calculated, and the dependence of the differential
cross section and the volume integral $J_{F}$ for the
Sn($^{3}$He,t)Sb reactions at E($^{3}$He)$=200$ MeV occurring by
the excitation of IAR on mass number A is examined. Our results
show that the calculated value for the isospin mixing in the
$^{100}$Sn isotope is in good agreement with Colo et al.'s
estimates $(4-5\%)$, and the obtained values for the volume
integral change within the error range of the value reported by
Fujiwara et al. (53$\pm$5 MeV fm$^{3}$). Moreover, it is concluded
that although the differential cross section of the isobar analog
resonance for the ($^{3}$He,t) reactions is not sensitive to
pairing correlations between nucleons, a considerable effect on
the isospin admixtures in $N\approx Z$ isotopes can be seen with
the presence of these correlations.
\end{abstract}

\section{Introduction}
With the development of the heavy mass ion accelerators
technology, many isotopes being proton rich within the range of
A=80-100 have been established. The studies on such isotopes, far
away from the beta stability valley, are very useful in
understanding many nuclear aspects. The investigation on the
isospin admixture effect in the ground state of these proton rich
nuclei can be considered as one of the possible candidate studies.
It is well known that the isospin admixture effects of the nuclear
states are very important in the estimates of the effective vector
coupling constants based on Fermi transitions, and in the
description of the energies and the widths of the analog states
and the isospin multiplets [1-4].

The isospin mixing is basically caused by the Coulomb potential,
or more accurately by that part of Coulomb potential which changes
over the nuclear volume. Since the very small symmetry energy
tries to minimize the difference in the proton and the neutron
systems, Coulomb forces are more dominant in nuclei for which the
proton and the neutron numbers are close to each other. Therefore,
the isospin mixing can take large values in the ground state of
the proton rich nuclei.

The physics of the exotic nuclei, characterized by the very
unusual ratios of the neutrons and the protons, can be considered
as a test for the already well-established models of the nuclear
structure that are used to describe the stable systems. Since the
parameters and the interactions used in the usual shell-model
calculations are determined in order to reproduce the properties
of the known nuclei, they may not always be appropriate for use in
the calculation of the nuclei approaching to the drip lines.

The isospin admixtures in the nuclear ground states have been
calculated in various models. The two liquid hydrodynamic model is
used to estimate the energies of the collective isovector monopole
excitation with an isospin of $T= T_{0}+1$ [5]. The admixture of
this excited state to the ground state with an isospin of $T=
T_{0}$, caused by the Coulomb potential, turned out to be small
$\approx(0.1-0.3)\%$ for all stable nuclei with A$>40$.
Quantitative estimates performed by using the shell model [6-8]
are approximately an order of magnitude larger than the estimate
of Bohr and Mottelson. There are several effects which may cause
such a difference in the mentioned estimates. First, the shell
model calculations were performed in the particle-hole
approximation and used a limited number of configurations. Second,
the residual isovector interaction was either neglected, or it was
included in the Tamm-Dankoff approximation. Third, the residual
interaction was not related to the shell-model potential in the
self consistent way. The residual isovector forces in separable
form are derived for a given form of the shell-model potential
[9]. The derivation is based on the isotopic invariance of the
nuclear forces. The form of the interaction strength parameters is
related to the potential given by the self-consistency relations.
It has been shown that the $T= T_{0}+1$ isospin admixture in the
ground state of the parent nucleus can be determined by the sum of
the square of the nuclear matrix elements for beta decay
$(N+1,Z-1)$ [9]. The absolute values of the admixtures, however,
are in some cases several times larger than the hydrodynamic
estimate. Theoretical calculations for the isospin mixing in
proton rich nuclei between A$=80-100$ have been performed [10-20].
The Hartree-Fock calculations for the very proton-rich nuclei in
the region around A$=80-100$ predict the isospin mixing in an
order of $3-5\%$ [13]. With the inclusion of the random phase
approximation in the calculations, the isospin mixing increased by
an amount of $15-20\%$ [10]. These estimates are still 2-3 times
larger than the Bohr and Mottelson ones [5]. The pairing
correlations between nucleons were not considered in all studies
mentioned above. However, it has been stated in Ref. [10] that
these correlations could be of primary importance for medium and
heavy nuclei.

In the present study, based on the spherical single-particle wave
functions and energies with the pairing and the residual isovector
interaction treated in QRPA, the isospin admixtures in the ground
states of the $^{100-124}$Sn isotopes and the differential cross
sections for the charge-exchange reactions ($^{3}$He,t) at
E($^{3}$He)$=200$ MeV occurring by the excitation of the IAR
states in the $^{112-124}$Sb were investigated. In the
calculations, considering the restoration of the isotopic
invariance for the nuclear part of the total Hamiltonian [9,21]
the effective interaction strength $(\gamma_{\rho})$ in the
quasiparticle space has been obtained in such a way that it is
self consistent with the Woods-Saxon form of the shell model
potential. Thus, this method makes our model free of any
adjustable parameters.

\section{Hamiltonian}
Let us consider a system of nucleons in a spherical symmetric
average field interacting via pairing forces. In this case, the
corresponding single quasiparticle Hamiltonian of the system is
given by
\begin{equation}
\widehat{H}_{sqp}=\sum\limits_{\tau ,j,m}\varepsilon _{j}^{(\tau
)}\alpha _{jm}^{\dagger }(\tau )\alpha _{jm}(\tau ),\ \ \ \ \ \ \
\ \ (\tau =n,p)~,
\end{equation}
where the $\varepsilon _{j}^{(\tau )}$  is the single
quasiparticles energy of the nucleons with  angular momentum $j$,
and the $\alpha _{jm}^{\dagger }(\tau )(\alpha _{jm}(\tau ))$ is
the quasiparticle creation (annihilation) operator.  The shell
model potential contains the isovector terms in nuclei having
different proton and neutron numbers in addition to the Coulomb
potential and proton-neutron mass difference  which breaks the
isotopic invariance (if the proton-neutron mass difference is
neglected)
\begin{equation}
\left[ \widehat{H}_{sqp}-V_{c},\widehat{T}^{\rho }\right] \neq 0~,
\end{equation}
where $V_{c}$ is the Coulomb potential given by the following
expression:
\begin{equation}
    V_{c}=\sum\limits_{i=1}^{A}v_{c}(r_{i})\left(
\frac{1}{2}-t_{z}^{i}\right),~~t_{z}^{i}= \left\{
\begin{array}{ll}
     ~ 1/2, & \hbox{for neutrons;} \\
    -1/2,& \hbox{for protons}~, \\
\end{array}
\right.
\end{equation}
with the radial part of the Coulomb potential
\begin{equation}
v_{c}(r)=\frac{e^{2}(Z-1)}{Z}{\displaystyle\int \frac{\rho
_{p}(r')}{|\overrightarrow{r}-\overrightarrow{r}'|}d\overrightarrow{r}'}~.
\end{equation}
Here, $\rho_{p}(r')$\ is to the proton density distribution in
ground state. The isospin operators $\widehat{T}^{\rho}\ $ are
defined in the following way:
\begin{equation}
\widehat{T}^{\rho }=\frac{1}{2}\left[ \widehat{T}_{+}+\rho
\widehat{T}_{-} \right] =\left\{
\begin{array}{ll}
    \widehat{T}_{x}, & \hbox{$\rho =+1$;} \\
    i\widehat{T}_{y}, & \hbox{$\rho =-1$.} \\
\end{array}%
\right. ,~~\widehat{T}_{\pm
}=\sum\limits_{k=1}^{A}\widehat{t}_{\pm }^{k}~,
\end{equation}
\noindent where $\widehat{t}_{\pm }^{k}= \widehat{t}_{x}^{k}\pm
i\widehat{t}_{y}^{k}$ are the raising and the lowering isospin
operators. The broken symmetry should be restored by the residual
interaction $\widehat{h}$ which obey the following equation [9,21]
\begin{equation}
\left[ \widehat{H}_{sqp}+\widehat{h}-V_{c},\widehat{T}^{\rho
}\right] =0~,
\end{equation}
where $\widehat{h}$ is defined as [22]
\begin{equation}
\widehat{h}=\sum\limits_{\rho =\pm 1}\frac{1}{4\gamma _{\rho
}}\left[ \widehat{H}_{sqp}-V_{c},\widehat{T}^{\rho }\right]
^{\dagger }~\left[ \widehat{H}_{sqp}-V_{c},\widehat{T}^{\rho
}\right]~.
\end{equation}
$\gamma _{\rho }$ is an average of double commutator in the ground
state,
\begin{equation}
\gamma _{\rho }=\frac{\rho }{2}\left\langle 0\left\vert \left[
\left[ \widehat{H}_{sqp}-V_{c},\widehat{T}^{\rho }\right]
,\widehat{T}^{\rho } \right] \right\vert 0\right\rangle~.
\end{equation}
Such a form of the residual interaction allows us to treat the
Coulomb mixing effects of the isospin simply and
self-consistently. Let us note that the derivation of Eq. (7) uses
only one assumption, namely the separability of the residual
interactions. Thus, the restoration of the isotopic invariance for
the nuclear part of the Hamiltonian is satisfied, and the total
Hamiltonian operator can be written in the form of
\begin{equation}
\widehat{H}=\widehat{H}_{sqp}+\widehat{h}.
\end{equation}
\section{Isobar Analog States}
In this section, we shall consider the isobaric $0^{+}$
excitations in odd-odd nuclei generated from the correlated ground
state of the parent even-even nucleus by the charge-exchange
forces and use the eigenstates of the single quasiparticle
Hamiltonian $\widehat{H}_{sqp}$ as a basis. The basis set for the
neutron-proton quasiparticle pair creation
$\widehat{A}_{j_{n}j_{p}}^{\dagger}$ and  annihilation
$\widehat{A}_{j_{n}j_{p}}$ operators is defined as :
\begin{equation}
\widehat{A}_{j_{n}j_{p}}^{\dagger }=\frac{1}{\sqrt{2j_{n}+1}}
\sum\limits_{m}(-1)^{j_{n}-m}\alpha _{j_{p}m}^{\dagger }\alpha
_{j_{n},-m}^{\dagger }\ ,
\end{equation}
$$
 \widehat{A}_{j_{n}j_{p}}=\left(
\widehat{A} _{j_{n}j_{p}}^{\dagger }\right) ^{\dagger}~.
$$
The bosonic commutation relations of these operators in
quasi-boson approximation [10] are given by
\begin{equation}
\left[ \widehat{A}_{j_{n}j_{p}},\widehat{A}_{j_{n {\acute{}}
}j_{p{\acute{}} }}^{\dagger }\right] \approx \delta _{j_{n}j_{n
{\acute{}} }}\delta _{j_{p}j_{p {\acute{}} }}~,\ \ \left[
\widehat{A}_{j_{n}j_{p}},\widehat{A}_{j_{n {\acute{}} }j_{p
{\acute{}} }}\right] =0~.
\end{equation}
In the quasiparticle space, the effective interaction
$\widehat{h}$ and the average of the double commutator $\gamma
_{\rho }$ can be written in the form of
\begin{equation}
\widehat{h}=-\sum\limits_{j_{n},j_{p},j_{n {\acute{}} },j_{p
{\acute{}} },\rho }\frac{1}{4\chi _{\rho }}E_{j_{n}j_{p}}^{\rho
}E_{j_{n{\acute{}} }j_{p {\acute{}} }}^{\rho }\left(
\widehat{A}_{j_{n}j_{p}}-\rho \widehat{A} _{j_{n}j_{p}}^{\dagger
}\right)\left( \widehat{A}_{j_{n {\acute{}} }j_{p {\acute{}}
}}^{\dagger }-\rho \widehat{A}_{j_{n {\acute{}} }j_{p {\acute{}}
}}\right)~,
\end{equation}
\[
\chi _{\rho }=-\gamma _{\rho
}=\sum\limits_{j_{n},j_{p}}b_{j_{n}j_{p}}^{\rho
}E_{j_{n}j_{p}}^{\rho }~,~b_{j_{n}j_{p}}^{\rho
}=\frac{1}{2}(\overline{b}_{j_{n}j_{p}}+\rho b_{j_{n}j_{p}})~,
\]
with
$$
E_{j_{n}j_{p}}^{\rho }=\left\{ \varepsilon
_{j_{n}j_{p}}b_{j_{n}j_{p}}^{\rho }+\frac{1}{2}\left(
\overline{\varphi }_{j_{n}j_{p}}-\rho \varphi _{j_{n}j_{p}}\right)
\right\}~,
$$
$$
 \varepsilon _{j_{n}j_{p}}=\varepsilon
_{j_{n}}+\varepsilon _{j_{p}}~,\nonumber\\
$$
$$
b_{j_{n}j_{p}}=\sqrt{2j_{p}+1}\left\langle j_{p}\left\Vert
\widehat{t}_{-}\right\Vert j_{n}\right\rangle
u_{j_{p}}v_{j_{n}}~,\nonumber\\
$$
$$
\overline{b}_{j_{n}j_{p}}=\sqrt{2j_{p}+1}\left\langle
j_{p}\left\Vert \widehat{t} _{-}\right\Vert j_{n}\right\rangle
u_{j_{n}}v_{j_{p}}~,\nonumber\\
$$
$$
\varphi_{j_{n}j_{p}}=\sqrt{2j_{p}+1}\left\langle j_{p}\left\Vert
\widehat{t} _{-}v_{c}(r)\right\Vert j_{n}\right\rangle
u_{j_{p}}v_{j_{n}}~, \nonumber\\
$$
\begin{equation}
\overline{ \varphi }_{j_{n}j_{p}}=\sqrt{2j_{p}+1}\left\langle
j_{p}\left\Vert \widehat{t }_{-}v_{c}(r)\right\Vert
j_{n}\right\rangle u_{j_{n}}v_{j_{p}}~,\\
\end{equation}
$$
\varepsilon _{j_{\tau }}=\left[ C_{\tau }^{2}+(E_{j_{\tau
}}-\lambda _{\tau })^{2}\right] ^{1/2}~,\nonumber\\
$$
$$
u_{j_{\tau }}=\left[ \frac{1}{2}\left( 1+\frac{ E_{j_{\tau
}}-\lambda _{\tau }}{\varepsilon _{j_{\tau }}}\right) \right]
^{1/2}~,\nonumber\\
$$
$$
v_{j_{\tau }}=\left[ \frac{1}{2}\left( 1-\frac{E_{j_{\tau
}}-\lambda _{\tau }}{\varepsilon _{j_{\tau }}}\right) \right]
^{1/2}~,\nonumber
$$
where $C_{\tau }$,$\ E_{j_{\tau }}$ and $\lambda _{\tau }$\
correspond to the correlation function, the single particle
energies for nucleons and the chemical potential, respectively.
$v_{j_{\tau }}(u_{j_{\tau }})\ $ is the occupation (unoccupation)
amplitude. In QRPA, the collective $0^{+}$ states, generated by
the effective interaction $\widehat{h}$, are accepted as
one-phonon excitations and can be described as follows:
\begin{equation}
|\psi _{i}\rangle =\widehat{Q}_{i}^{\dag }|0\rangle
=\sum\limits_{j_{n},j_{p}}\left( r_{j_{n}j_{p}}^{i}\widehat{A}%
_{j_{n}j_{p}}^{\dagger
}-s_{j_{n}j_{p}}^{i}\widehat{A}_{j_{n}j_{p}}\right)|0\rangle,
\end{equation}
where $r_{j_{n}j_{p}}^{i}$, and $s_{j_{n}j_{p}}^{i}$ are the
amplitudes for the neutron-proton quasiparticle pair,
$\widehat{Q}_{i}^{\dag }$ is the phonon creation operator, and
$|0\rangle $\ is the phonon vacuum corresponding to the ground
state of the even-even nucleus, i.e.
\begin{equation}
\widehat{Q}_{i}|0\rangle =0.
\end{equation}
Assuming that the phonon operators obey the commutation relations
given below
\begin{equation}
\langle 0\left\vert \left[ \widehat{Q}_{i},\widehat{Q}_{j}^{\dag
}\right] \right\vert 0\rangle =\delta _{ij~\ },~\ \langle
0\left\vert \left[ \widehat{ Q}_{i},\widehat{Q}_{j}\right]
\right\vert 0\rangle =0,
\end{equation}
we obtain the following orthonormalization condition for the
amplitudes:
\begin{equation}
\sum\limits_{j_{n},j_{p}}\left(
r_{j_{n}j_{p}}^{i}r_{j_{n}j_{p}}^{i'}-s_{j_{n}j_{p}}^{i}s_{j_{n}j_{p}}^{i'}\right)
=\delta _{ii'}.
\end{equation}
The eigenvalues and the eigenfunctions of the restored Hamiltonian
(9) can be obtained by solving the equation of motion in QRPA
\begin{equation}
\left[ \widehat{H}_{sqp}+\widehat{h},\widehat{Q}_{i}^{\dag
}\right] |0\rangle =\omega _{i}\widehat{Q}_{i}^{\dag }|0\rangle .
\end{equation}
Here the $\omega _{i}$'s are the energies of the isobaric $0^{+}$
states. Employing the conventional procedure of QRPA, we obtain
the dispersion equation for the excitation energy of the isobaric
$0^{+}$ states as follows:
%\begin{equation}
\begin{align}
\left[ \chi _{+1}-\sum\limits_{j_{n},j_{p}}\frac{\varepsilon
_{j_{n}j_{p}}\left( E_{j_{n}j_{p}}^{+1}\right) ^{2}}{\varepsilon
_{j_{n}j_{p}}^{2}-\omega _{i}^{2}}\right] &\left[\chi
_{-1}-\sum\limits_{j_{n},j_{p}}\frac{\varepsilon
_{j_{n}j_{p}}\left( E_{j_{n}j_{p}}^{-1}\right) ^{2}}{\varepsilon
_{j_{n}j_{p}}^{2}-\omega _{i}^{2}}\right]\nonumber\\ &-\omega
_{i}^{2}\sum\limits_{j_{n},j_{p}}\left[ \frac{\varepsilon
_{j_{n}j_{p}}E_{j_{n}j_{p}}^{+1}E_{j_{n}j_{p}}^{-1}}{\varepsilon
_{j_{n}j_{p}}^{2}-\omega _{i}^{2}}\right] ^{2}=0.
\end{align}
The amplitudes for the neutron-proton quasiparticle pair can then
be expressed analytically in the following form:
$$
r_{j_{n}j_{p}}^{i}=\frac{1}{\sqrt{Z(\omega _{i})}}\frac{%
E_{j_{n}j_{p}}^{+1}+L(\omega _{i})E_{j_{n}j_{p}}^{-1}}{\varepsilon
_{j_{n}j_{p}}-\omega _{i}} ,\\
$$
\begin{equation}
s_{j_{n}j_{p}}^{i}=\frac{1}{\sqrt{Z(\omega _{i})}
}\frac{E_{j_{n}j_{p}}^{+1}-L(\omega
_{i})E_{j_{n}j_{p}}^{-1}}{\varepsilon _{j_{n}j_{p}}+\omega _{i}} ,
\end{equation}
with
$$
L(\omega _{i})=\frac{\chi
_{+1}-\sum\limits_{j_{n},j_{p}}\frac{\varepsilon
_{j_{n}j_{p}}\left( E_{j_{n}j_{p}}^{+1}\right) ^{2}}{\varepsilon
_{j_{n}j_{p}}^{2}-\omega _{i}^{2}}}{\omega
_{i}\sum\limits_{j_{n},j_{p}} \frac{\varepsilon
_{j_{n}j_{p}}E_{j_{n}j_{p}}^{+1}E_{j_{n}j_{p}}^{-1}}{ \varepsilon
_{j_{n}j_{p}}^{2}-\omega _{i}^{2}}} .
$$
The quantity $Z(\omega _{i})$ is determined from the normalization
condition given in Eq. (17).

An analog state is contained in the solutions of the Eq.(19). This
is easily seen in case of the constant Coulomb potential
\begin{equation}
\left\langle j_{p}\left\Vert \widehat{t}_{-}^{}v_{c}(r)\right\Vert
j_{n}\right\rangle =\Delta E_{C}\left\langle j_{p}\left\Vert
\widehat{t} _{-}\right\Vert j_{n}\right\rangle ,~\Delta E_{C}={\rm
constant.}
\end{equation}
The equation (19) now contains the solution $\omega _{k}=\Delta
E_{C}$, corresponding to the average energy of the single
quasi-particle Coulomb shift of the nuclei $(N,Z)$ and
$(N-1,Z+1)$. From Eqs. (14), (17) and (20), it follows that
\begin{equation}
\widehat{Q}_{k}^{\dag }\mid 0\rangle _{\omega _{k}=\Delta
E_{C}}=\frac{1}{ \sqrt{2T_{0}}}\widehat{T}_{-}\mid 0\rangle ,
\end{equation}
i.e., this solution describes the IAR state. Because Eq. (16) is
fulfilled, the isospin is exactly conserved in all states. This
latter fact is a natural consequence of the isospin invariance of
the self-consistent nuclear Hamiltonian.

\section{Fermi Beta Transitions}
It has been shown in Ref. [9] that if the pairing correlations
between nucleons are not considered, two independent isobaric
excited $0^{+}$ states occur from the ground state of parent
nucleus: the $T_{z} =T_{0}-1$ states including Isobar Analog
Resonance (IAR) in the nucleus with the number $(N-1,Z+1)$ (the
isotopic spin of the parent nucleus is assumed as $T=T_{0}$), and
the $T_{z} =T_{0}+1$ excited states in the nucleus with the number
$(N+1,Z-1)$. It must also be noted that both branches in question
are dependent on the mother nucleus due to the Fermi transition
and the matrix elements for the $\beta^{\pm}$ transitions obeying
Fermi sum rule.

Let us now mention how the above situation changes with the
presence of the pairing forces. First of all, there will be no two
independent isobaric $0^{+}$ excited states, and these states will
occur in both nuclei with the number $(N-1,Z+1)$ and $(N+1,Z-1)$
due to the violation of particle number conservation. Secondly,
the total probability of the $\beta$ transition from both nuclei
$(N+1,Z-1)$ and $(N-1,Z+1)$ to the parent nucleus (N,Z) increases
in such a way that the Fermi sum rule is fulfilled.

The isobaric $0^{+}$ states in the neighbor odd-odd nuclei
$(N-1,Z+1$, and $N+1,Z-1)$ is characterized by the Fermi
transition matrix elements between these states and the ground
state of the neighbor even-even nuclei [4]. One could obtain the
following Fermi transition matrix elements by using the wave
functions in Eq.(14):

a) For the transitions $(N,Z)\Longrightarrow (N-1,Z+1)$,
\begin{equation}
M_{\beta ^{-}}^{i}=\langle 0\left\vert
\left[\widehat{Q}_{i},\widehat{T}_{-}\right] \right\vert 0\rangle
=\sum\limits_{j_{n},j_{p}}\left(
r_{j_{n}j_{p}}^{i}b_{j_{n}j_{p}}+s_{j_{n}j_{p}}^{i}\overline{b}
_{j_{n}j_{p}}\right) ,
\end{equation}

b) For the transitions $(N,Z)\Longrightarrow (N+1,Z-1)$,
\begin{equation}
M_{\beta ^{+}}^{i}=\langle 0\left\vert
\left[\widehat{Q}_{i},\widehat{T}_{+}\right] \right\vert 0\rangle
=\sum\limits_{j_{n},j_{p}}\left(
r_{j_{n}j_{p}}^{i}\overline{b}_{j_{n}j_{p}}+s_{j_{n}j_{p}}^{i}b_{j_{n}j_{p}}
\right) .
\end{equation}
It is possible to show that the transitions in question obey the
Fermi sum rule
\begin{equation}
\sum\limits_{i}\left\{ \left\vert M_{\beta ^{-}}^{i}\right\vert
^{2}-\left\vert M_{\beta ^{+}}^{i}\right\vert ^{2}\right\}
=2T_{0}=N-Z~.
\end{equation}
In case of the constant Coulomb potential (21), only the matrix
elements of the analog resonance $\beta$ decay is non-vanishing,
and $M_{\beta ^{-}}^{i}(\omega_{i}=\Delta E_{C})=\sqrt{2T_{0}}$.
This matrix element exhausts the full strength of the Fermi
transition. The separation of the IAR state appearing in Eqs.
(19-22) from the other states and the collection of all the
$\beta$ transition strength on this state are the evidence for the
conservation of the isotopic invariance for the nuclear part of
the shell model Hamiltonian.

\section{Isospin Structure of the Ground State for the Parent Nuclei}
When the isospin structure of the ground state for the nuclei
considered in the present study is investigated, it is observed
that the isospin impurity of the ground states and IAR are related
to the matrix elements of the $\beta^{\pm}$ transitions
($M_{\beta^{\pm}}$) [5]. Expanding the ground state wave function
in terms of the pure isospin components $|T,T_{z}\rangle$, we
obtain
\begin{equation}
|0\rangle =a|T_{0},T_{0}\rangle +b|T_{0}+1,T_{0}\rangle ,~\ \ \
a^{2}+b^{2}=1 .
\end{equation}
This expansion states that the ground state of the parent nucleus
will contain only the $T_{0}+1$ isospin admixtures caused by the
isovector Coulomb potential. The expectation value of the square
of the isospin in the ground state of the parent nucleus can be
expressed as
\begin{equation}
\left\langle 0\left\vert \widehat{T}^{2}\right\vert 0\right\rangle
=T_{0}(T_{0}+1)+\sum\limits_{i}\left\vert M_{\beta
^{+}}^{i}\right\vert ^{2}.
\end{equation}
On the other hand, the following expression for the same quantity
can be obtained from Eq. (26)
\begin{equation}
\left\langle 0\left\vert \widehat{T}^{2}\right\vert 0\right\rangle
=T_{0}(T_{0}+1)+2b^{2}(T_{0}+1) .
\end{equation}
From Eqs. (27) and (28), it follows that the $T_{0}+1$ isospin
admixture in the ground state of the parent nucleus is determined
by the sum of the squares of the beta decay matrix elements from
the isobaric states of the nucleus $(N+1,Z-1)$:
\begin{equation}
b^{2}=\left[ 2(T_{0}+1)\right] ^{-1}\sum\limits_{i}\left\vert
M_{\beta ^{+}}^{i}\right\vert ^{2} .
\end{equation}
This result obtained in Ref [9] is given over the sum of the
$\beta^{+}$ transition matrix elements from the parent nucleus to
the neighbor odd-odd (N+1, Z-1) nuclei. However, in the similar
studies, the $T_{0}+1$ isospin admixture is usually determined by
the Coulomb mixing of the ground state with the isovector monopole
excited states in the same even-even nucleus [13,23].

\section{Differential Cross Sections of $^{112-124}$Sn($^{3}$He,t)$^{112-124}$Sb Reactions}

If the nucleon motion of the $^{3}$He beam is much faster than the
relative Fermi motion of the nucleons in $^{3}$He, the
($^{3}$He,t) reaction is expected to become simple like the (p,n)
reaction at intermediate energies. The experiment in Ref. [24]
indicates that the complex contribution from the Fermi motion of
the nucleons in $^{3}$He and in the triton is negligible in the
charge-exchange reactions at the bombarding energy around $150$
MeV/nucleon.

The differential cross sections of zero degrees for the excitation
of the IAR can be written as [25,26]
\begin{equation}
\left( \frac{d\sigma }{d\Omega }\right) _{F}(q\approx 0,~\theta
=0)=\left( \frac{\mu }{\pi \hbar ^{2}}\right) ^{2}\left(
\frac{k_{f}}{k_{i}}\right) N_{F}J_{F}^{2}B(F),
\end{equation}
where $J_{F}$\ is the volume integral of the central part of the
effective interaction, $N_{F}$\ is the distortion factor which may
be approximated by the function $exp(-xA^{1/3})$ [26], $\mu ~$ and
$k$ denote the reduced energy divided by $c^{2}$ and the wave
number in the center of mass system, respectively. The value of
$x$ is taken from Ref. [24] and $B(F)=\left\vert M_{\beta
^{-}}^{i=IAR}\right\vert ^{2}$\ is the reduced matrix elements.

\section{Results and Discussions}
In this section, the isospin admixtures for the $^{100-124}$Sn
isotopes, the IAR energies in the $^{112-124}$Sb isotopes, and the
differential cross section for the Sn$(^{3}$He,t)Sb reactions at
E($^{3}$He)$=200$ MeV occurring by the excitation of the IAR state
were numerically calculated by considering the pairing interaction
between nucleons. In the calculations, the Woods-Saxon Potential
with Chepurnov parametrization [27] was used, and the correlation
function ($C_{n}=12/\sqrt{A}$) was chosen in accordance with Ref.
[23]. The basis used in our calculation contain all neutron-proton
transitions which change the radial quantum number n by $\Delta
n=0,1,2,3$. The left-hand side of the sum-rule in Eq.(25)
containing the overlap integrals $\langle n\|p\rangle$ is
fulfilled with the approximately $\sim 1\%$ accuracy.

Our results for the $T_{0}+1$ isospin admixtures in the ground
state of the $^{100-124}$Sn isotopes with (solid line) and without
(dotted line) pairing correlations between nucleons have been
presented in Fig. 1. As seen from Fig. 1, the admixture amplitude
($b^{2}$) decreases with the increasing N-Z value in both cases
since according to Pauli principle the number of the $\beta^{+}$
transitions with $\Delta n\neq 0$ which makes a considerable
contribution to the sum in Eq. (27) will decrease as the neutron
number increases. It can also be seen obviously from this figure
that the pairing correlations between nucleons lead to shift the
isospin admixture values ($b^{2}$) to the lower ones. This effect
is more pronounced in case of the small N-Z values although it
vanishes for the large N-Z values. This is an expected result
because these pairing correlations are more dominant when the
isovector potential is small. The nearness  of the isospin
admixture values for the $^{100}$Sn isotope are due to the weak
effect of the pairing correlations in the closed shell nuclei.
\begin{figure}[h]
\includegraphics[width=4.7in]{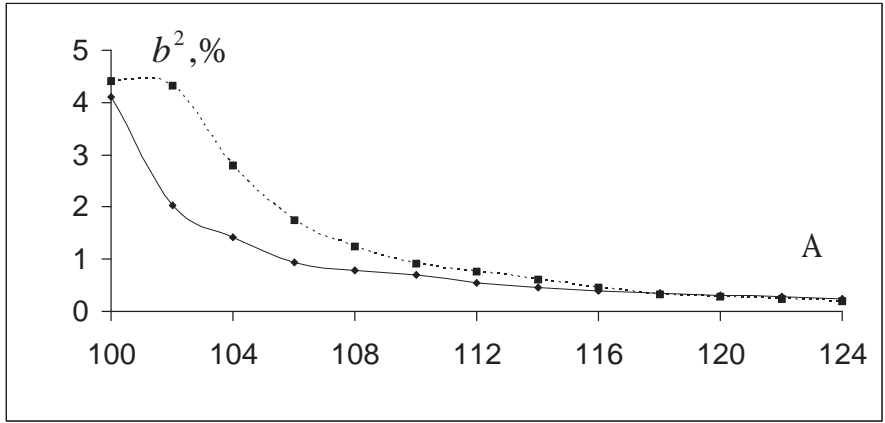} \\
\begin{center}
     % Give the correct figure height in cm
\caption{Dependence of the contribution of the $T_{0}+1$
isomultiplet states to the ground of even-even Sn isotopes  on
mass number A. The solid and dashed lines correspond to the cases
with and without the pairing forces, respectively}
\label{fig:1}       % Give a unique label
\end{center}
\end{figure}

In Fig. 2, our model results (solid line) for the quantity $b^{2}$
in $^{100-124}$Sn isotopes have been compared with the
hydrodynamic model results [23] (dashed line) and the values
calculated by using the formula given in Eq. (11) of Ref. [13]
(dotted line).
\begin{figure}[h]
\includegraphics[width=4.7in]{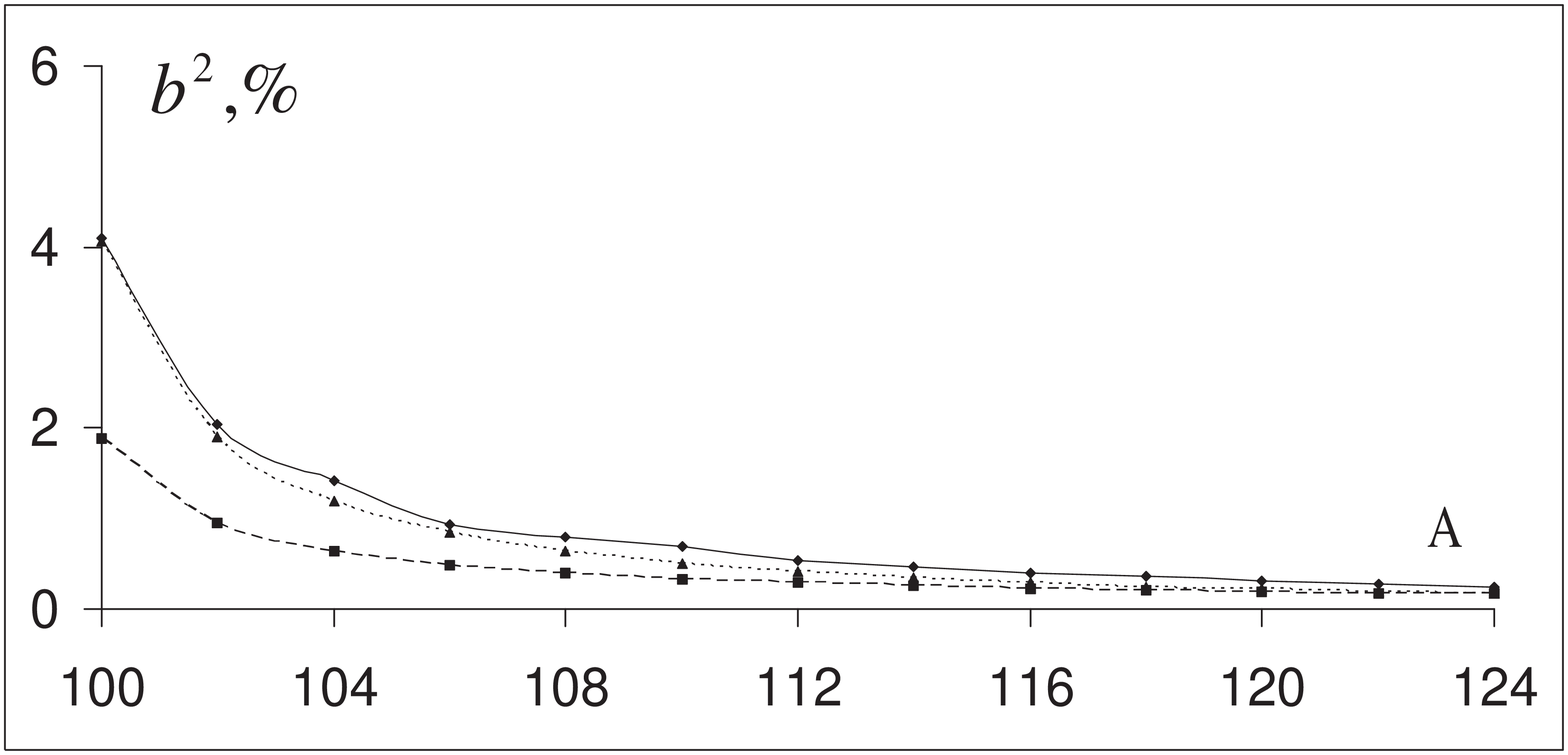} \\
\begin{center}
\caption{Dependence of  the calculated values for the quantity
$b^{2}$ in Sn isotopes with respect to the different models on
mass number A. The solid and dashed lines correspond to our model
and the hydrodynamic model results. The values of Ref. [13] are
given in the dotted lines}
\label{fig:2}       % Give a unique label
\end{center}
\end{figure}
Our results are closer to those given in Ref. [13] and two times
larger than the hydrodynamic models for the $^{100}$Sn isotope.
The difference between the results of the different models
diminishes with an increasing mass number A. The calculated
results shown in Fig. 2 are also numerically given in Table 1.
\begin{table}[h]
\begin{center}
\caption{The numerical results for the quantity $b^{2}$ obtained
by the different models and our model in $^{100-124}$Sn isotopes}
\label{tab:1}
\begin{tabular}{cccc}
\hline\noalign{\smallskip} A & Bohr and  & Calculation results &
This work \\
  & Mottelson [23] & of Ref. [13]&    \\ \hline
100 & 1.885 & 4.067 & 4.100 \\ \hline 102 & 0.955 & 1.908 & 2.037
\\ \hline
104 & 0.645 & 1.197 & 1.421 \\ \hline 106 & 0.490 & 0.847 & 0.930
\\ \hline
108 & 0.397 & 0.641 & 0.702 \\ \hline 110 & 0.334 & 0.506 & 0.691
\\ \hline
112 & 0.290 & 0.411 & 0.543 \\ \hline 114 & 0.257 & 0.343 & 0.462
\\ \hline
116 & 0.231 & 0.290 & 0.398 \\ \hline 118 & 0.211 & 0.249 & 0.356
\\ \hline
120 & 0.194 & 0.217 & 0.306 \\ \hline 122 & 0.179 & 0.190 & 0.276
\\ \hline
124 & 0.167 & 0.168 & 0.248 \\ \hline
\end{tabular}
\end{center}
\end{table}

The dependence of the IAR energies for the $^{100-124}$Sb isotopes
on mass number A has been shown in Fig. 3. The IAR energies were
calculated from the ground state of the Sb isotopes. As seen from
Fig. 3, the pairing forces try to shift the IAR energies to the
lower values in all Sb isotopes studied here. The corresponding
energy shift for the $^{104-122}$Sb isotopes changes from 529 keV
to 53 keV. The pairing correlations between nucleons play an
important role on the IAR energy shift (up to 0.5 MeV) in the
light isotopes as it was in the case of the isospin admixture
($b^{2}$). In Table 2, the calculated values of the IAR energies
for $^{112-124}$Sb isotopes based on different models have been
compared with the experimental ones. Our results are in agreement
to the experimental ones compared with the results in Ref [28].
\begin{figure}[h]
\includegraphics[width=4.7in]{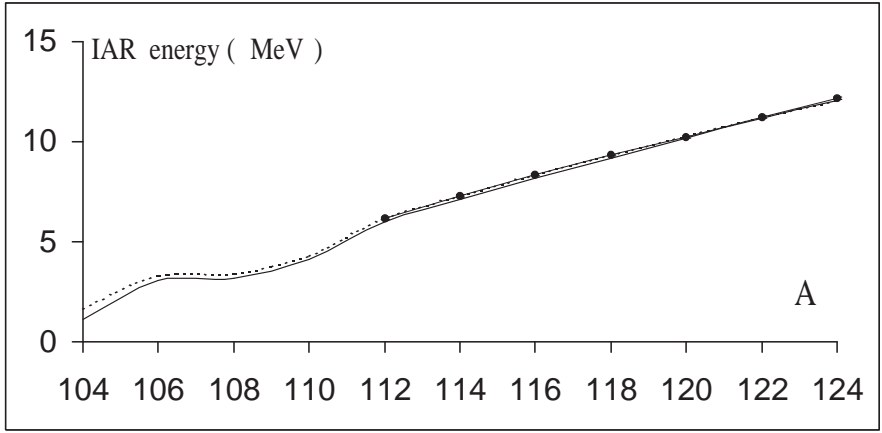} \\
\begin{center}
\caption{Dependence of the IAR energies on mass number A in the Sb
isotopes. The solid and dotted lines correspond to the cases with
and without the pairing forces, and points show the experimental
values [29]}
\label{fig:3}       % Give a unique label
\end{center}
\end{figure}

The differential cross-section for the
$^{112-124}$Sn$(^{3}$He,t)$^{112-124}$Sb reactions at
E($^{3}$He)$=200$ MeV occurring by the excitation of the IAR
states has been calculated by using the formula in Eq. (30). In
this calculation, the value for the central volume integral is
taken as $J_{F}=53$ MeV fm$^{3}$ [24] and the distortion factor
$N_{F}$ is considered as $N_{F}=exp(-0.7A^{1/3})$. Our model
results with (solid line) and without (dotted line) pairing
correlations have been compared with the experimental values [29]
(dashed line) in Fig. 4. From the corresponding curves, it is
obvious that the contribution of the pairing forces to the
differential cross-section is small, and the theoretical values of
the differential cross-section for the
$^{112-124}$Sn$(^{3}$He,t)$^{112-124}$Sb isotopes are still larger
than the experimental ones.

\begin{table}[h]
\begin{center}
\caption{The IAR energies in the $^{112-124}$Sb isotopes (in MeV)}
\label{tab:2}
\begin{tabular}{cccc}
\hline\noalign{\smallskip}
A & Our Calculations & Rodin and & Experiment [29]\\
  &  & Urin [28]
\\ \hline
112 & 5.994 & 5.6 & 6.16$\pm$0.03 \\ \hline 114 & 7.123 & 6.8 &
7.28$\pm$0.03
\\ \hline
116 & 8.191 & 7.8 & 8.36$\pm$0.03 \\ \hline 118 & 9.193 & 8.8 &
9.33$\pm$0.03
\\ \hline
120 & 10.149 & 9.7 & 10.24$\pm$0.03 \\ \hline 122 & 11.203 & 11.2
& 11.24$\pm$0.03
\\ \hline
124 & 12.087 & 12.2 & 12.19$\pm$0.03 \\ \hline
\end{tabular}
\end{center}
\end{table}

The value of $J_{F}=53\pm 5$ MeV fm$^{3}$ for the volume integral
has been obtained in Ref. [24] by considering  the ($^{3}$He,t)
reaction at E($^{3}$He)$=450$ MeV. In our study, the volume
integral $J_{F}$ has been calculated by using the experimental
value [29] of the differential cross-section for the ($^{3}$He,t)
reaction at E($^{3}$He)$=200$ MeV. The calculation results have
been depicted in Fig. 5. The value of the volume integral $J_{F}$
has a tendency to decrease with some fluctuations as the mass
number A increases. Here, the effect of pairing forces are also
weak. Our calculated $J_{F}$ values for the mass number region,
$A=112-124$, are changing within the range of the value of $53\pm
5$ MeV fm$^{3}$ given in Ref. [24].
\begin{figure}[h]
\includegraphics[width=4.7in]{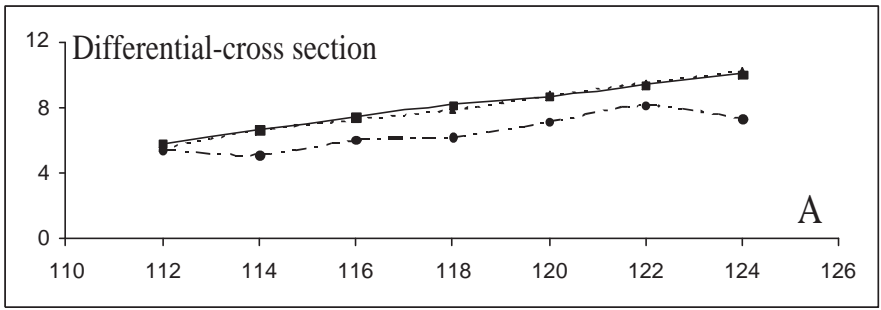} \\
\begin{center}
\caption{Dependence of  the differential cross-section,
$\frac{d\sigma}{d\Omega}$ (in mb/sr), for the Sn($^{3}$He,t)Sb
reactions at E($^{3}$He)$=200$ MeV occurring by the excitation of
the IAR state on mass number A. The solid (dotted) and dashed
lines correspond to the cases with (without) the pairing
correlations, and the experimental values [29]}
\label{fig:4}       % Give a unique label
\end{center}
\end{figure}

In summary, the effect of the pairing interaction and the
isovector correlation  between nucleons on the properties of the
IAR state and the $T_{0}+1$ isospin admixture in even-even
isotopes has been investigated. The form and the strength
parameters of the interaction has been related to the shell model
potential by the self consistency relations. These relations make
our model free of any adjustable parameters. As a result of our
calculations, it has been observed that the $T_{0}+1$ isospin
admixture in $N\approx Z$ isotopes is sensitive to the pairing
interactions although the differential cross-section of the IAR
state for the ($^{3}$He,t) reactions is not sensitive to these
correlations. The value of $4.236\%$ obtained for the $T_{0}+1$
isospin admixture $b^{2}$ in $^{100}$Sn isotope shows a good
agreement with the value of $4-5\%$ [13]. Moreover, it is found
that the calculated values for the volume integral based on the
experimental differential cross-section values [29] of the
($^{3}$He,t) reactions at E($^{3}$He)$=200$ MeV were in agreement
with the value of  $53\pm 5$ MeV fm$^{3}$ [24].
\begin{figure}[h]
\includegraphics[width=4.7in]{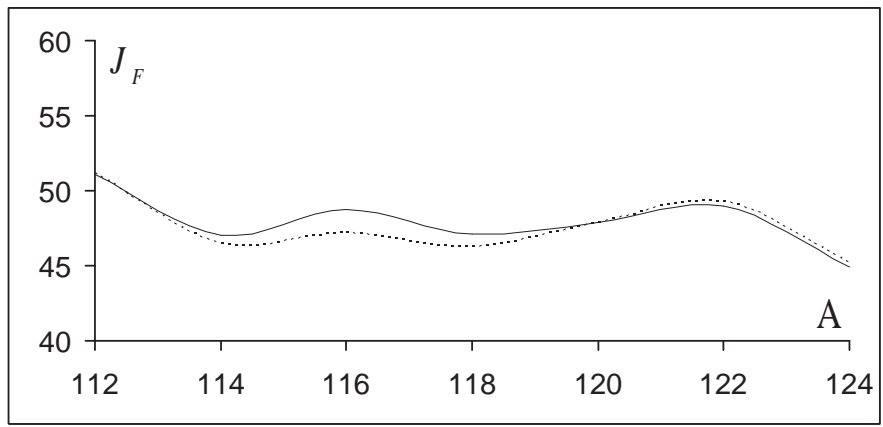} \\
\begin{center}
\caption{Dependence of the volume integral $J_{F}$ (in MeV
fm$^{3}$) obtained from the Sn($^{3}$He,t)Sb reaction occurring by
the excitation of the IAR state at E($^{3}$He)$=200$ MeV,
$\theta=0^{\circ}$ on mass number A. The solid and dotted lines
are the cases with and without the pairing forces.}
\label{fig:5}       % Give a unique label
\end{center}
\end{figure}

\thanks{\textbf{Acknowledgement:}} We are very grateful to Professor A. A.
Kuliev for his contributions to our study.

\end{document}